\documentclass[showpacs,superscriptaddress,amsmath,amssymb,nofootinbib,twocolumn,floatfix]{revtex4}

\usepackage{setspace}
\usepackage{graphicx}  % Include figure files
\usepackage{dcolumn}   % Align table columns on decimal point
\usepackage{bm}        % bold math
\usepackage{rotating}
\usepackage{lipsum}
\usepackage{hyperref}
\usepackage{multirow}
\newcommand{\msun}{\rm {M\textsubscript{\(\odot\)}}}
\newcommand{\deff}{D_{\rm{eff}}}
\newcommand{\dlumi}{D_{\rm{L}}}
\begin{document}

%\preprint{CERN-PH-EP-2011-162}
%\preprint{Submitted to Physical Review D}

% ======================
% TITLE AND ABSTRACT
% ======================

\title{Searches for Compact Binary Coalescence Events using Neural Networks in LIGO/Virgo Second Observation Period}

\author{A. Men\'endez-V\'azquez}
\affiliation{Institut de F\'\i sica  d'Altes Energies (IFAE), Barcelona Institute of Science and Technology, E-08193 Barcelona, Spain}
\author{M. Kolstein}
\affiliation{Institut de F\'\i sica  d'Altes Energies (IFAE), Barcelona Institute of Science and Technology, E-08193 Barcelona, Spain}
\author{M. Mart\'\i nez}
\affiliation{Institut de F\'\i sica  d'Altes Energies (IFAE), Barcelona Institute of Science and Technology, E-08193 Barcelona, Spain}
\affiliation{Catalan Institution for Research and Advanced Studies (ICREA), E-08010 Barcelona, Spain}
\author{Ll. M. Mir}
\affiliation{Institut de F\'\i sica  d'Altes Energies (IFAE), Barcelona Institute of Science and Technology, E-08193 Barcelona, Spain}
\
\date{\today}

% ===================
% ABSTRACT
% ===================

\begin{abstract}

We present results on the search for the coalescence of compact binary mergers using convolutional neural networks  and 
the  LIGO/Virgo data, 
corresponding to the O2 observation period. Two-dimensional images in time and frequency are used as input, 
and two sets of neural networks are trained separately for low  mass (0.2 -  2.0 $\msun$) 
and high mass (25 - 100 $\msun$) compact binary coalescence events. 
We explored neural networks trained with input information from a single or a pair of interferometers, indicating that the use of information 
from pairs leads to an improved performance. A scan over the full O2 data set using the 
convolutional neural networks for detection
demonstrates that the performance is compatible with 
that from canonical pipelines using matched filtering techniques, opening the possibility for an online implementation in future 
observation periods. No additional   
events with significant signal-to-noise ratio are found in the O2 data.

\end{abstract}

%\keywords{Suggested keywords} % Use showkeys class option if keyword
                               % display desired

\maketitle

% ==============
%  INTRODUCTION
% ==============

\section{Introduction}
\label{sec:intro}
Since the detection of a gravitational wave (GW) in 2015 \cite{FirstGWDet} a new era of gravitational wave astronomy has opened. 
This was confirmed with the detection of up to 11 events at the end of the second observation run (O2)~\cite{GWTC1}. Additional $39$ events were recently reported corresponding to  the first part of the third observation run (O3), 
for a total of 50 events~\cite{O3Catalog}. 
All the detected events to date are compatible with being originated by compact binary coalescence (CBC) of black holes (BH) or neutron stars (NS). 
The LIGO and Virgo Collaborations use matched-filtering techniques to extract the events from the much larger background 
(for a comprehensive review of the experimental techniques see Ref.~\cite{LIGOScientific:2019hgc}).  
The presence of a distinct chirp-like shape in the CBC events, 
when represented in spectrograms showing the signal in  frequency-time domain, 
makes the use of a convolutional neural network (CNN)  a valid alternative suitable for GW detection.  
In addition, the implementation online of a CNN pipeline for GW detection in the near future could translate into a significat 
reduction of the computation loads typically associated with the matched-filtering technique.  Once properly trained with a reduced dataset, 
the CNN has the potential to provide a fast and computationally efficient feedback on the presence of GW signals in the rest of the data.   

A machine learning approach for GW astronomy has been studied severely along the 
years~\cite{cuoco2020enhancing,PhysRevD.101.083006,Kim_2015}. 
In particular, 
different CNNs have been previously studied for the detection of GW 
events ~\cite{PhysRevLett.120.141103,george2017deep,Gebhard_2019,George_2018}
and in distinguishing between families of glitches~\cite{Razzano_2018,Biswas_2013,Cavaglia_2019}.
In this paper, 
we focus on the detection of CBC events with either very low ($0.2 - 2 \ \msun$) 
or very high  ($25 - 100 \ \msun$) mass ranges, 
and explore a CNN  based on a ResNet50 architecture~\cite{DBLP:journals/corr/HeZRS15} 
that has shown to give good results in image classification.

% =============
% DATA ANALYSIS
% =============
\section{Data preparation}
\label{sc:dp}

The study uses the O2 open data~\cite{O2OpenData} from LIGO-Livingston (L1), LIGO-Hanford (H1) and Virgo (V1) interferometers with 4096 Hz sampling rate.  
After applying quality requirements, the samples have a total duration of 154.0, 157.8,  and 20.8 days for L1, H1 and V1, respectively.  
A fraction of the data, 7.4 days in each interferometer, concentrated in the initial segments of the full O2 dataset and representative of the interferometer performance, 
is used for constructing background and background plus injected signal images for the purposes of the 
CNN training. The resulting total number of images are enough for an adequate training of the network.  
It constitutes a small fraction (about  5$\%$) 
of  the L1 and H1 data set but it amounts to  36$\%$ of the V1 data. Special precaution was taken in the preparation of the background images  to 
avoid including any of the identified GWs events in O2, as collected in the O2 catalog.

Waveforms for GW signals are generated using the IMRPhenomPv2~\cite{PhenomPv3} model and combined with data segments from the different interferometers, after taking 
into account the proper relative orientations, times of arrival and antenna factors.  In the case of the low mass CBC regime,
masses in the range between  0.2 -  2.0 $\msun$  are considered for the two compact objects in the binary system, and the corresponding luminosity distance $\dlumi$ is
limited to nearby events in the range 1 - 50 Mpcs.  In the case of high mass regime,  
signals with masses in the range between 25 and 100 $\msun$ and $\dlumi$ in the range between 100 and 1000 Mpcs are considered. Other parameters related to the position in the sky and orientation of the source are taken as homogeneously distributed. This finally results in a  signal
grid with O(250000) different signals. The injected signals are limited to a fixed
maximum duration of  five seconds. The five-seconds window is computed backward from the merger
    time to remove low-amplitude monochromatic-like parts of the waveform
    and avoid confusing the network during training.  A low frequency threshold of 80 Hz (20 Hz) is  applied for  the low-mass (high-mass) signal grid, 
 in order to control the duration of the injected signal.
Finally, the signals are randomly placed within the five-seconds window. \\

\begin{figure}[htb]
\begin{center}
\mbox{
\includegraphics[width=0.495\textwidth]{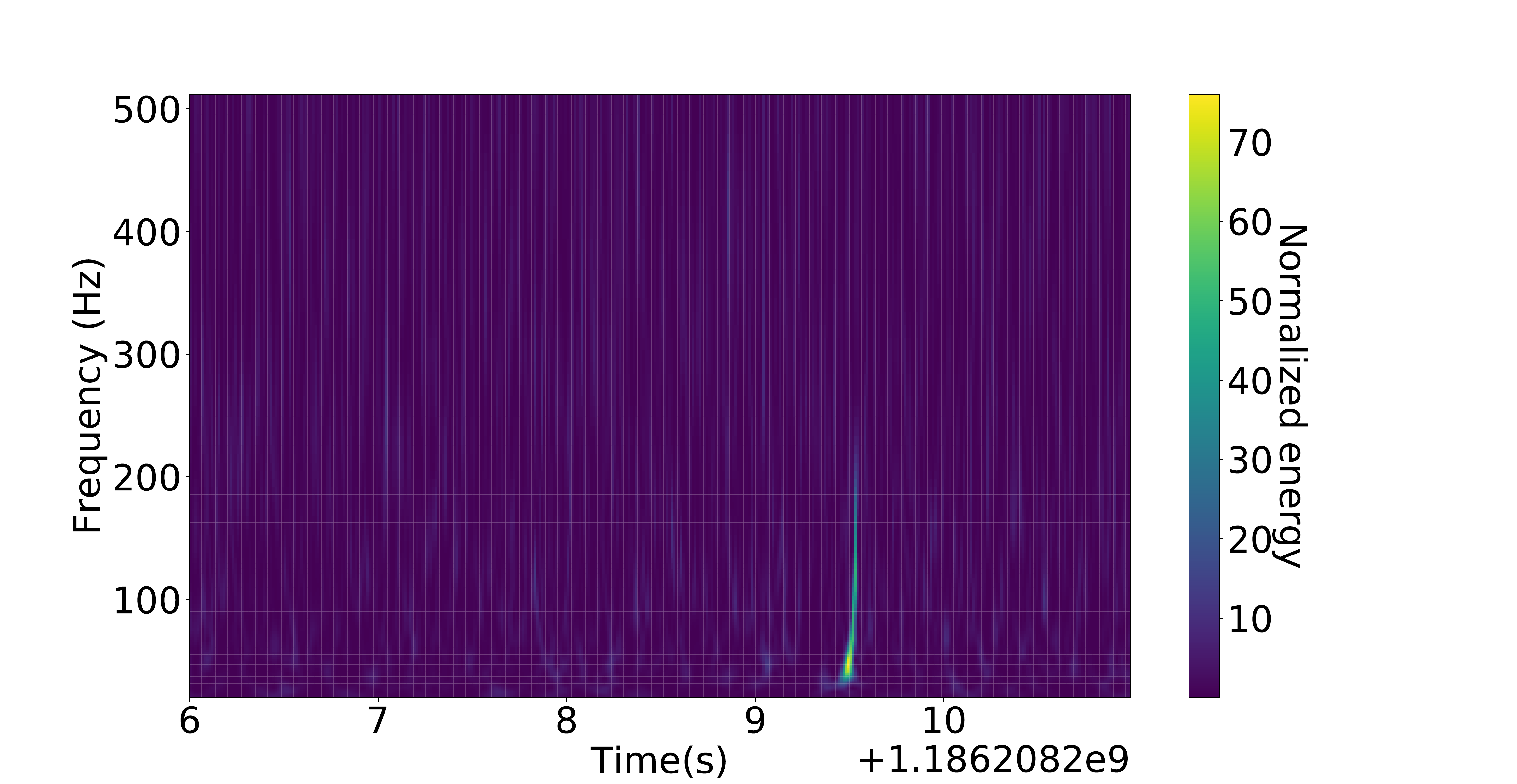}
}
\end{center}
\caption{\small
Example of a spectrogram  as a two-dimensional image in time versus frequency corresponding to a CBC  injected signal in the L1 interferometer with BH masses of 51~$\msun$   and 53~$\msun$ at a distance $\dlumi = 664$~Mpcs.  
}
\label{fig:spectro}
\end{figure}

Once the GW signals are injected  in the different L1, H1 and V1 background data segments, the data are processed. First, the data are whitened following the same prescription 
as in Ref.~\cite{LIGOScientific:2019hgc}. Two-dimensional arrays holding spectrogram data are then produced using $Q$-transforms~\cite{QTransf},    
in order to arrive to the desired images in terms of amplitude vs time vs frequency, with 400 bins in time and 100 bins in frequency. Figure~\ref{fig:spectro} presents an example of a spectrogram for a GW signal with BH masses of 51 $\msun$ and 53 $\msun$,  and $\dlumi = 664$~Mpcs, for which the GW signal is clearly observed. \\

In the low mass regime, and after taking into account the distance and the antenna factors for the detection, the strength of the GW signal is such that it often becomes invisible in the images and
constitutes a challenge for the CNN training. Hence,  to avoid a potential bias, we limit the signal injections to those events for which the effective distance $\deff$, defined as
\begin{equation}
\deff = \frac{\dlumi}{\sqrt{(1+\cos^2(l))^2 \ F^2_{+}/4 + \cos^2(l) \ F^2_{\times}}},    
\end{equation}
\noindent
is smaller than 60 Mpcs, 
where $l$ denotes the inclination angle,
and  $F_+$ and $F_\times$ are the antenna factors for the interferometer corresponding to the two GW polarizations.  When the  CNN receives pairs of signals from two different detectors as input (see section~\ref{sc:nn}), we require that at least 
one of them fulfills the $\deff\ < 60$~Mpcs requirement.   In the case of high mass range,  signals are loud enough and no requirements on $\deff$ are applied.

% =============
% NN definition and training
% =============
\section{Neural network definition and training}
\label{sc:nn}

We adopted, as nominal, a deep CNN {\it{ResNet-50}} with a 50-layer architecture, which demonstrates a good performance for image recognition.
As shown in Table~\ref{tab:arch}, the architecture
of the CNN follows closely that of the 50-layers layout of Ref.~\cite{DBLP:journals/corr/HeZRS15},  with modifications in the
last layers, for which average pooling and a fully connected dense layer (1-d fc)  with sigmoid activation is implemented.
%  as described in Ref.~\cite{DBLP:journals/corr/HeZRS15}, which demonstrates a good performance for image recognition. 
Alternatively, we explored the implementation of a CNN similar to that used in Ref.~\cite{Razzano_2018} for the 
detection and classification of noise in GW detectors.  The latter did not show a better performance. As already mentioned in Sec.~\ref{sc:dp}, 
two-dimensional images from different interferometers  
are input to the different CNNs, for which real data from LIGO/Virgo O2 observation period is employed to build the background and signal+background images.  A total of 128000 images per interferometer are used,  
evenly divided into background-only and background+signal. 
About 63$\%$ of the sample is devoted for training,  whereas a 7$\%$ and a 30$\%$ are used for validation and testing, respectively.   
Two separate CNNs are trained for the low-mass and the high-mass ranges.  
In the course of the CNN training,  it was observed that the presence of glitches in the data was not completely suppressed by the whitening process. This translates into 
large variations, image-by-image,  in the amplitude that in turn results into 
instabilities related to the batch normalization layers~\cite{BatchNorm}.
This was solved after renormalizing the contents in each image by its average in such a way that the contents in an image have 
an average equal to zero and a variance equal to one.

%                                                                                                                                                                                                                                           % ---  TABLE                                                                                                                                                                                                                                %                                                                                                                                                                                                                                            

\newcommand{\blocka}[2]{\multirow{3}{*}{\(\left[\begin{array}{c}\text{3$\times$3, #1}\\[-.1em] \text{3$\times$3, #1} \end{array}\right]\)$\times$#2}
}
\newcommand{\blockb}[3]{\multirow{3}{*}{\(\left[\begin{array}{c}\text{1$\times$1, #2}\\[-.1em] \text{3$\times$3, #2}\\[-.1em] \text{1$\times$1, #1}\end{array}\right]\)$\times$#3}
}
\renewcommand\arraystretch{1.1}
\begin{table}[htb]
\begin{center}
% \resizebox{0.7\linewidth}{!}{                                                                                                                                                                                                              
\footnotesize
\begin{tabular}{c|c|c}
\hline
layer name & output size & 50-layer  \\
\hline
conv1 & 112$\times$112 & \multicolumn{1}{c}{7$\times$7, 64, stride 2}\\
\hline
\multirow{4}{*}{conv2\_x} & \multirow{4}{*}{56$\times$56} & \multicolumn{1}{c}{3$\times$3 max pool, stride 2} \\
\cline{3-3}
  &  & \blockb{256}{64}{3} \\
  &  &  \\
  &  &  \\
\hline
\multirow{3}{*}{conv3\_x} &  \multirow{3}{*}{28$\times$28} & \blockb{512}{128}{4} \\
  &  &  \\
  &  &  \\
\hline
\multirow{3}{*}{conv4\_x} & \multirow{3}{*}{14$\times$14}  & \blockb{1024}{256}{6}\\
  &  &  \\
  &  &  \\
\hline
\multirow{3}{*}{conv5\_x} & \multirow{3}{*}{7$\times$7}  & \blockb{2048}{512}{3} \\
  &  &  \\
  &  &  \\
\hline
\multirow{3}{*}{} & 1$\times$1  & \multicolumn{1}{c}{global average pool, 1-d fc, sigmoid} \\
\hline
%\multicolumn{2}{l|}{FLOPs} & 3.8$\times10^9$ \\
%\hline
\multicolumn{3}{l}{Hyper parameters} \\ \hline
\multicolumn{2}{l}{Learning rate} & 0.001 \\
\multicolumn{2}{l}{Mini batch size} & 32 \\
\multicolumn{2}{l}{Maximum number of epochs} & 12 \\
\multicolumn{2}{l}{Optimizer} & Adam \\
\multicolumn{2}{l}{Loss function} & Binary-cross entropy \\ \hline
\end{tabular}
% }                                                                                                                                                                                                                                        
\end{center}
\vspace{-.5em}
\caption{
CNN architecture and the associated hyper parameters. Building blocks are shown in brackets, with the numbers of blocks stacked.
Downsampling is performed by conv3\_1, conv4\_1, and conv5\_1 with a stride of 2 (partially taken from~\cite{DBLP:journals/corr/HeZRS15}).
}
\label{tab:arch}
\vspace{-.5em}
\end{table}

\begin{figure}[htb]
\begin{center}
\mbox{
\includegraphics[width=0.495\textwidth]{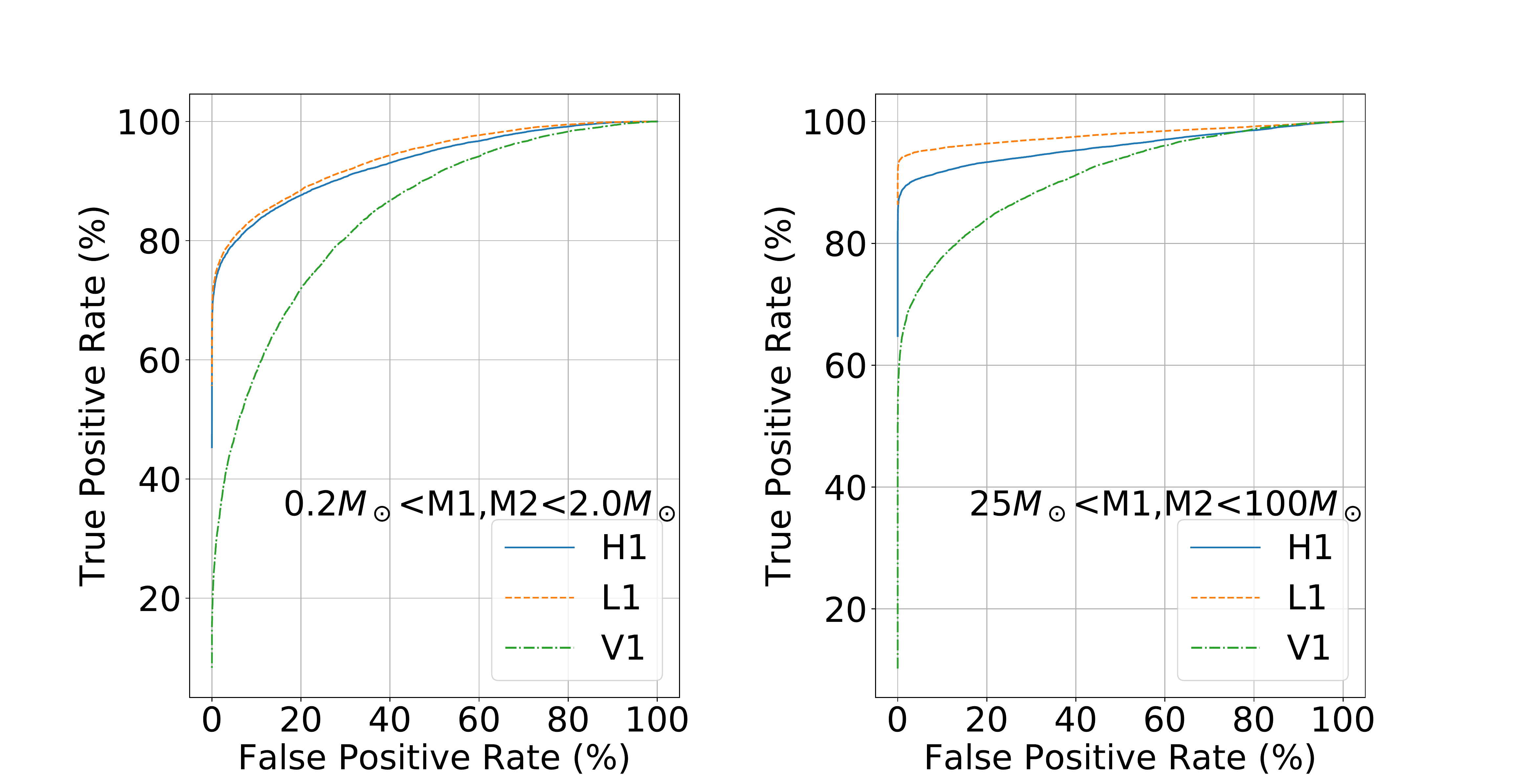}
}\\
\mbox{
\includegraphics[width=0.495\textwidth]{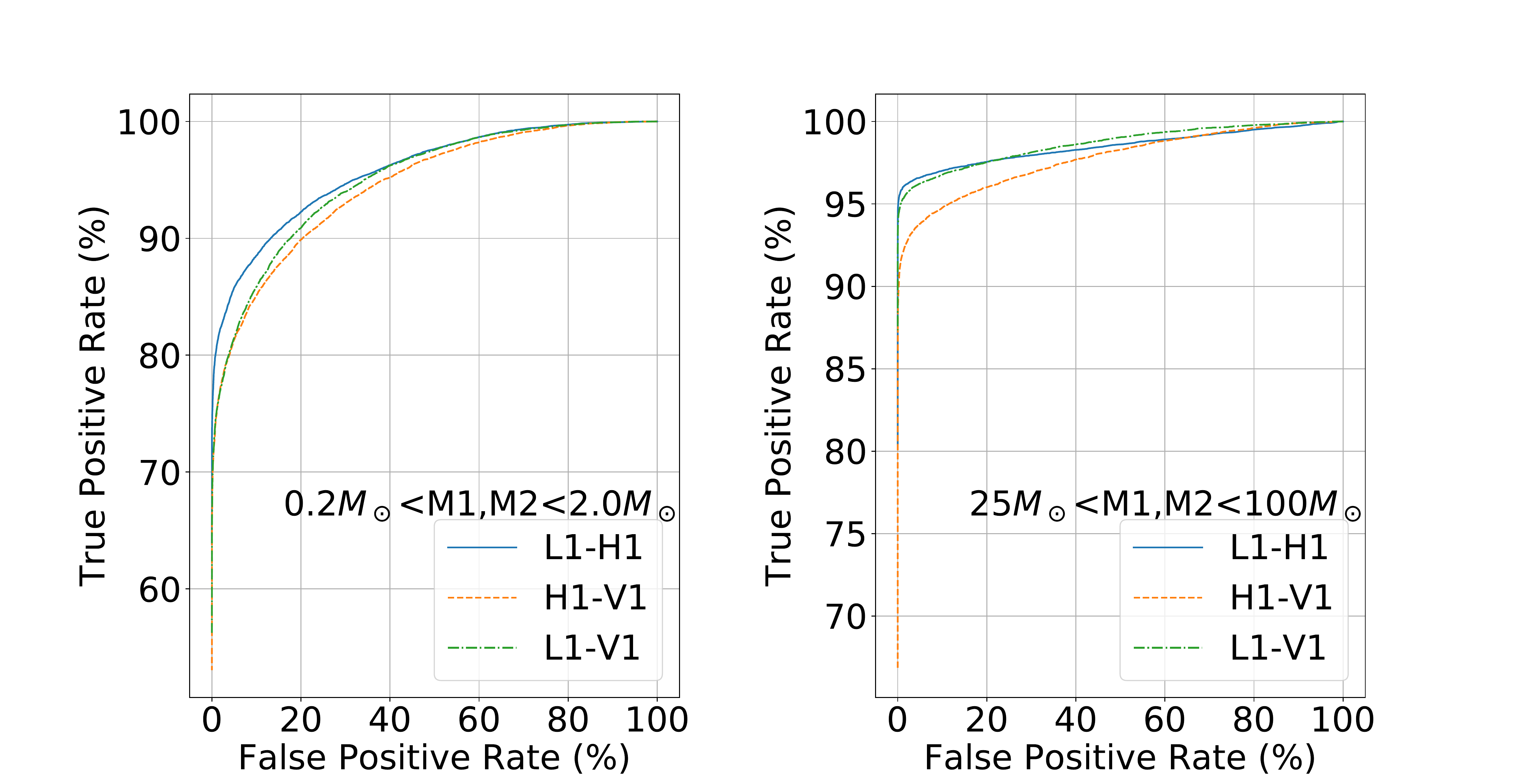}
}
\end{center}
\caption{\small
ROC curves  for the CNNs using  (top) single and (bottom) pairs  of interferometer inputs for (left) low mass and (right) high mass ranges.  
}
\label{fig:roc}
\end{figure}

We first explored training three separate  CNNs for L1, H1 an V1 data. 
For a number of epochs greater than five, the three CNNs reach stability. In the low-mass range and 
for the case of L1 and H1, a validation accuracy of the order of 88$\%$ is 
obtained. The V1 accuracy is smaller and a value around  75$\%$ is reached. 
Similarly, the validation loss  ranges 
from 0.2  in the case of L1 and 0.4 in the case of H1,   up to 0.7 in the case of V1.
In the high mass range, the CNNs present a better performance, to large extent attributed to the larger signal strain. 
The CNNs for L1, H1 and V1 show a validation accuracy of about 96$\%$, 92$\%$, and 82$\%$, and validation losses of 0.23, 0.03 and 0.60, respectively.
Altogether, these variations reflect the differences in sensitivity across interferometers. This is also     
clearly observed in Figure~\ref{fig:roc}-top,  showing the 
receiver operating characteristic (ROC)  curves for the separate CNNs,   
representing the true positive (TP) versus the false positive (FP) rates, where the differences 
in sensitivity between L1, H1 and V1 become evident.

\begin{figure}[htb]
\begin{center}
\mbox{
\includegraphics[width=0.495\textwidth]{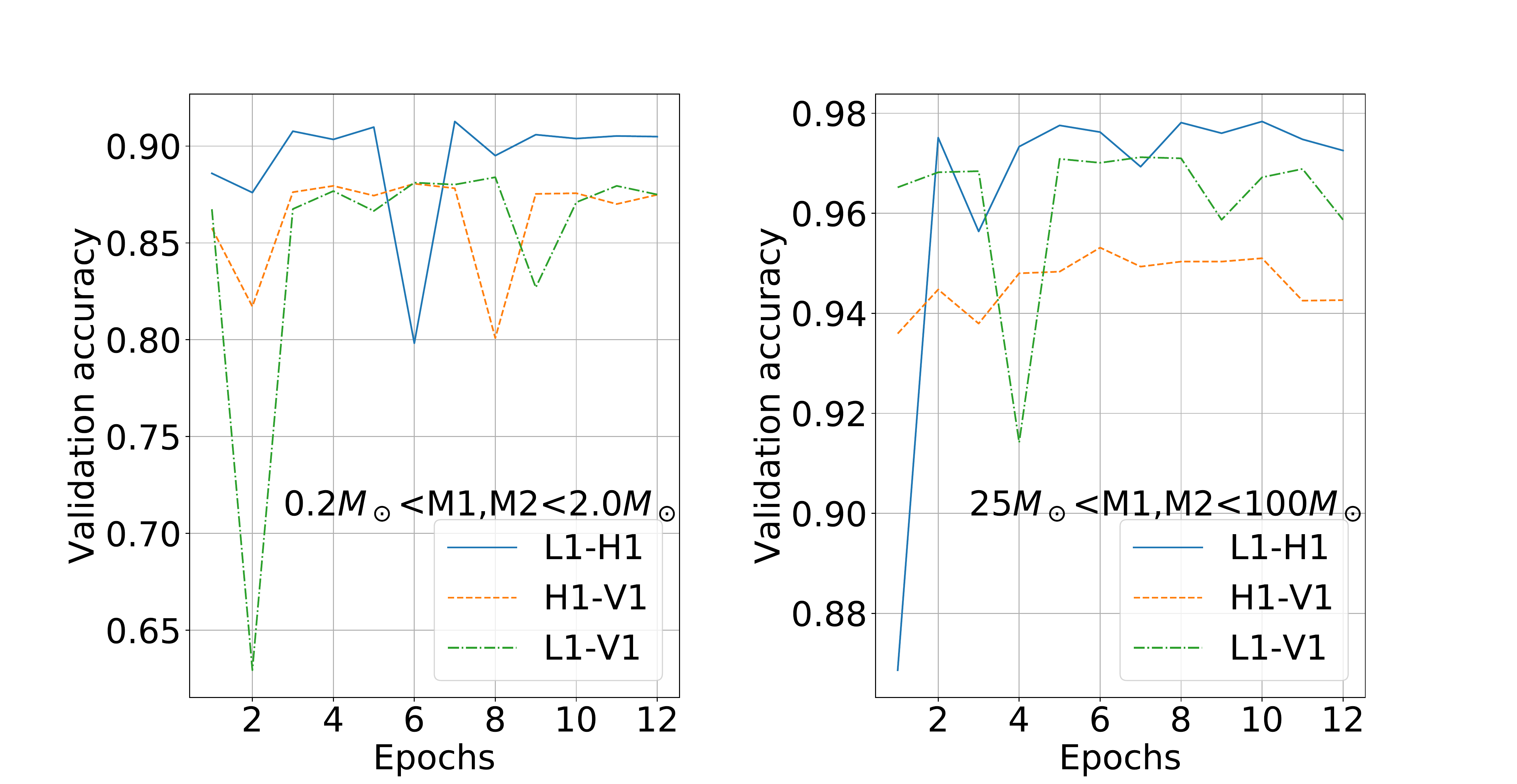}
}\\
\mbox{
\includegraphics[width=0.495\textwidth]{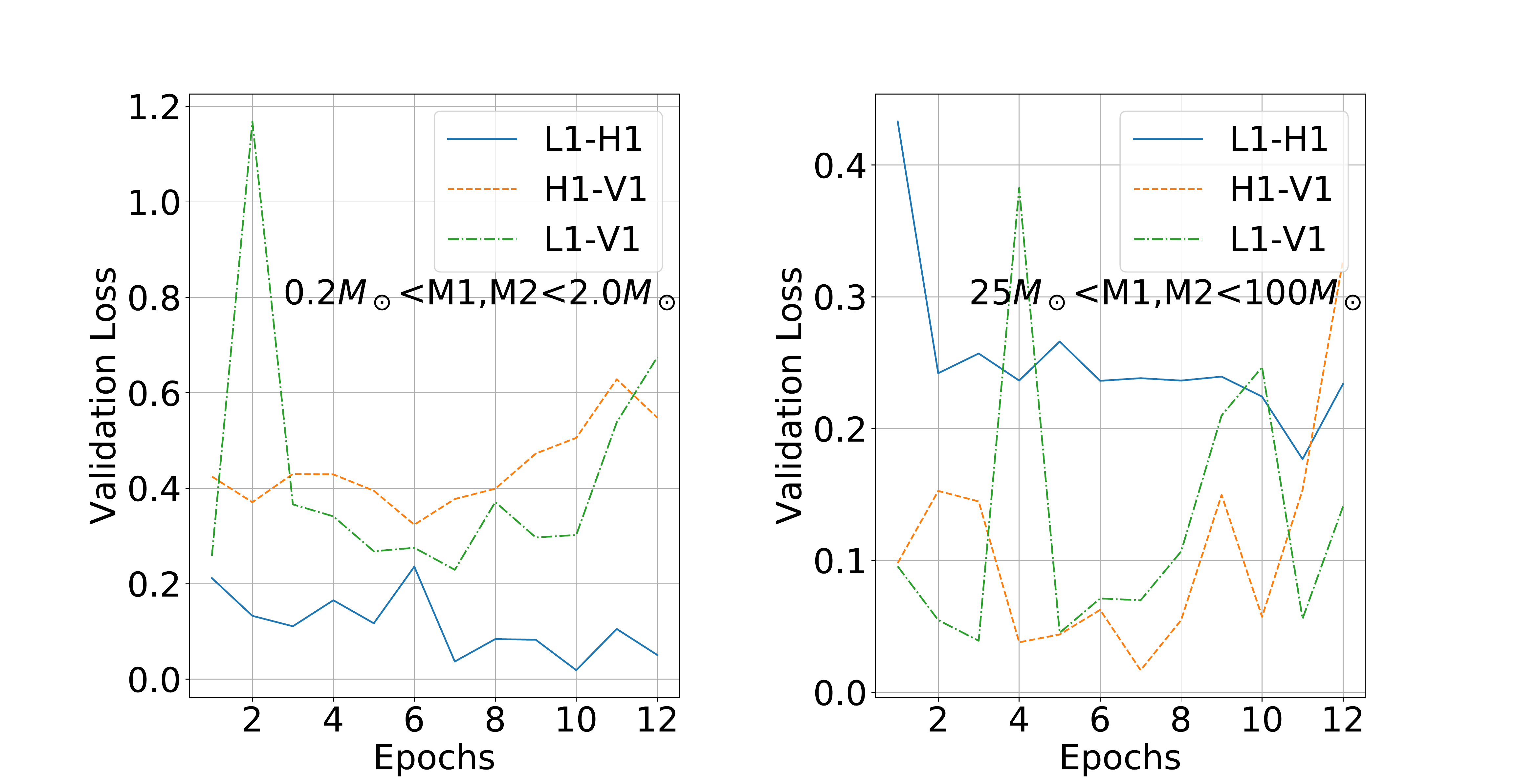}
}
\end{center}
\caption{\small
Validation accurary and validation loss as a function of the epochs for (left) the low mass range and (right) the high mass range, 
in the case of pairs of interferometers as input. 
}
\label{fig:val}
\end{figure}

The performance of the CNNs can be improved by including the information of pairs of interferometers during the training process. 
In this way, the CNN learns about the
correlations between images in two different channels when the signal is present. Given the limited duration of the V1 O2 data, the 
simultaneous use of the three interferometers as input was outside the scope of this study, but remains a natural extension of the work towards O3 and O4 observations runs. Figure~\ref{fig:val} shows, for the low and high mass ranges,  the evolution of the validation and loss accuracy as a function of epochs, demonstrating stability after about eight to ten epochs, 
with an accuracy of 91$\%$(97$\%$) and a loss below 0.1 (0.3) for the low (high) mass range.  
As expected, the inclusion of the V1 information in O2 did not translate into a significant
improvement,  whereas the L1-H1 combination leads to a slightly better performance of the CNN (see the  ROC curves in Figure~\ref{fig:roc}-bottom). 
 
\begin{figure}[htb]
\begin{center}
\mbox{
\includegraphics[width=0.245\textwidth]{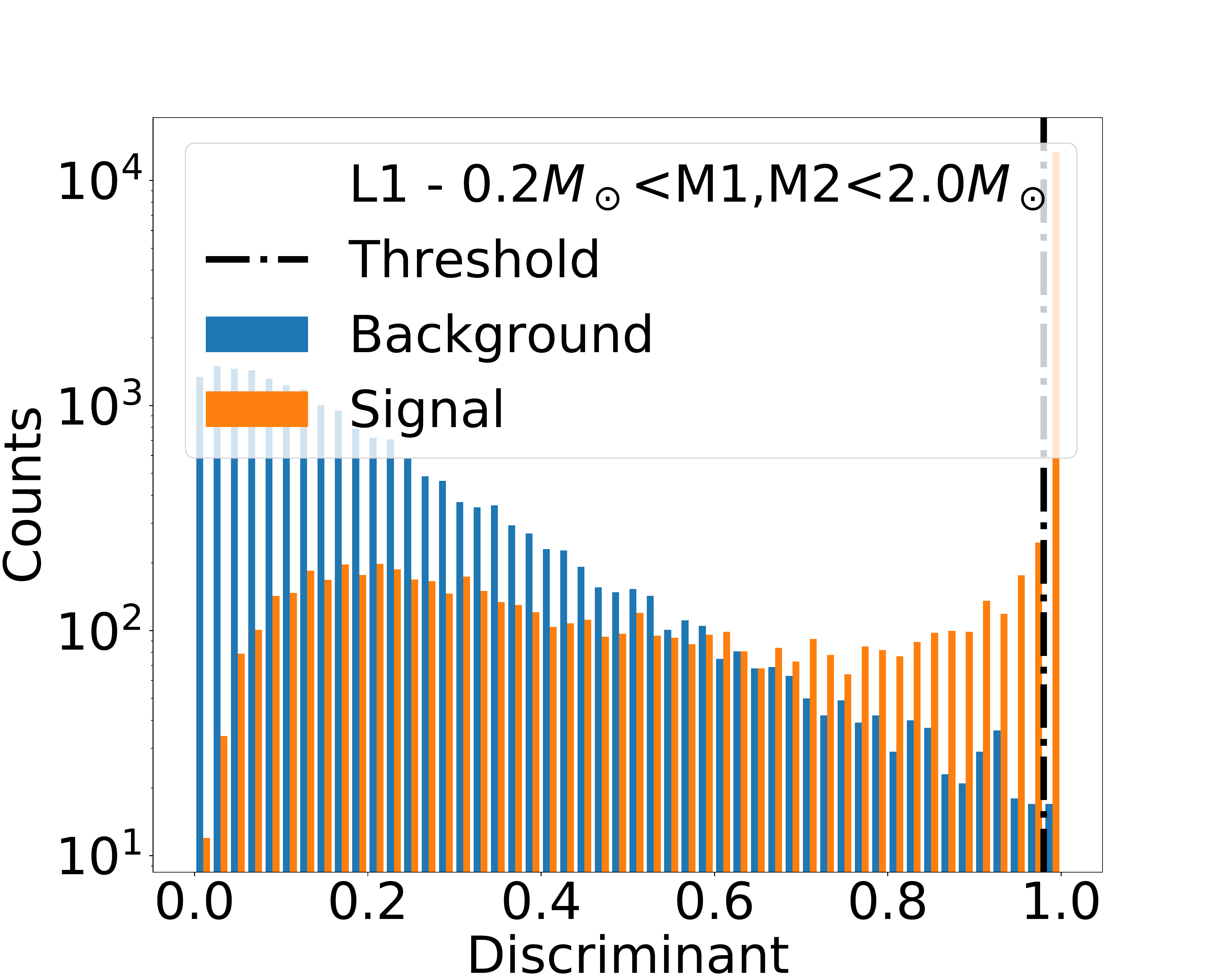}
\includegraphics[width=0.245\textwidth]{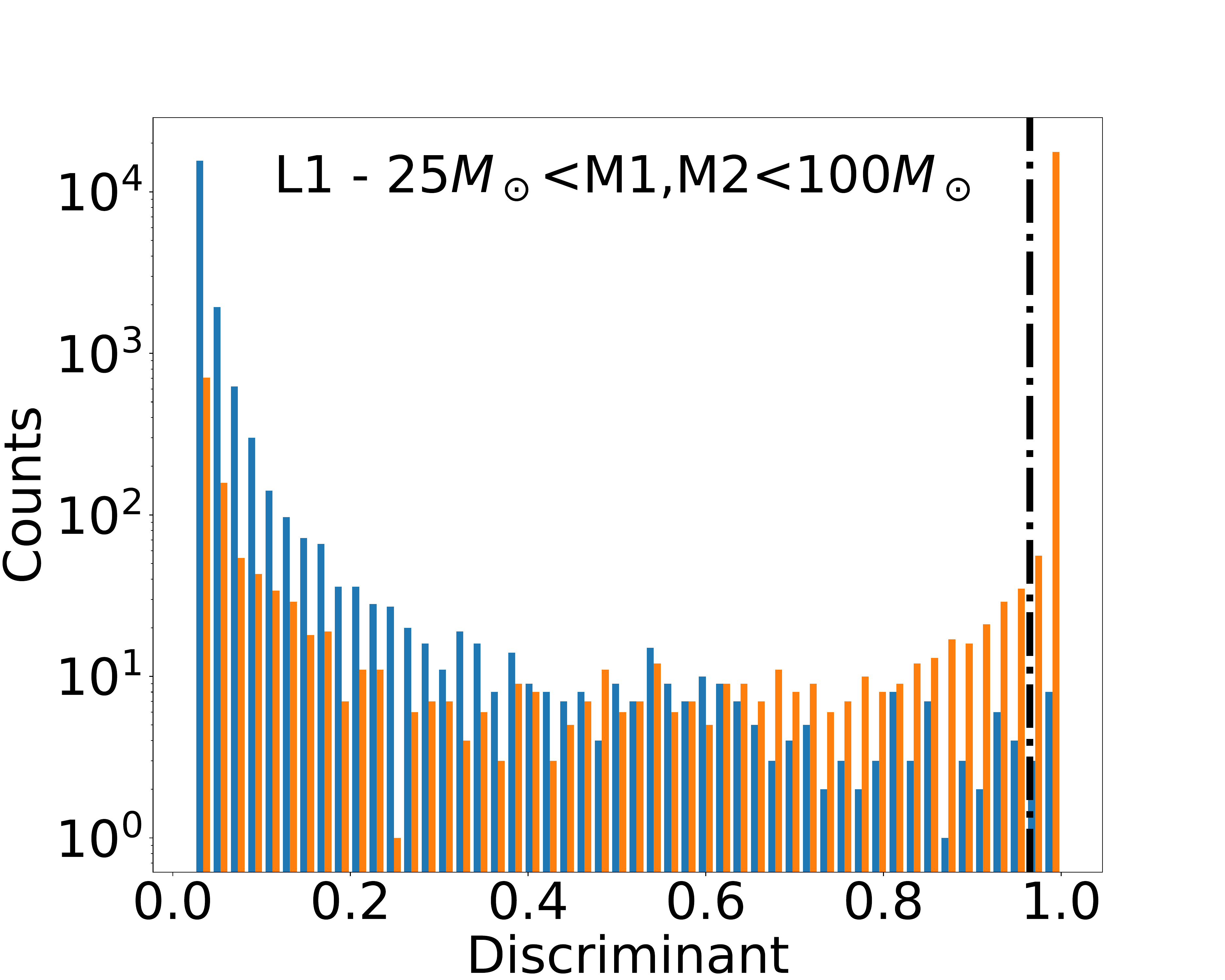} 
}
\mbox{
\includegraphics[width=0.245\textwidth]{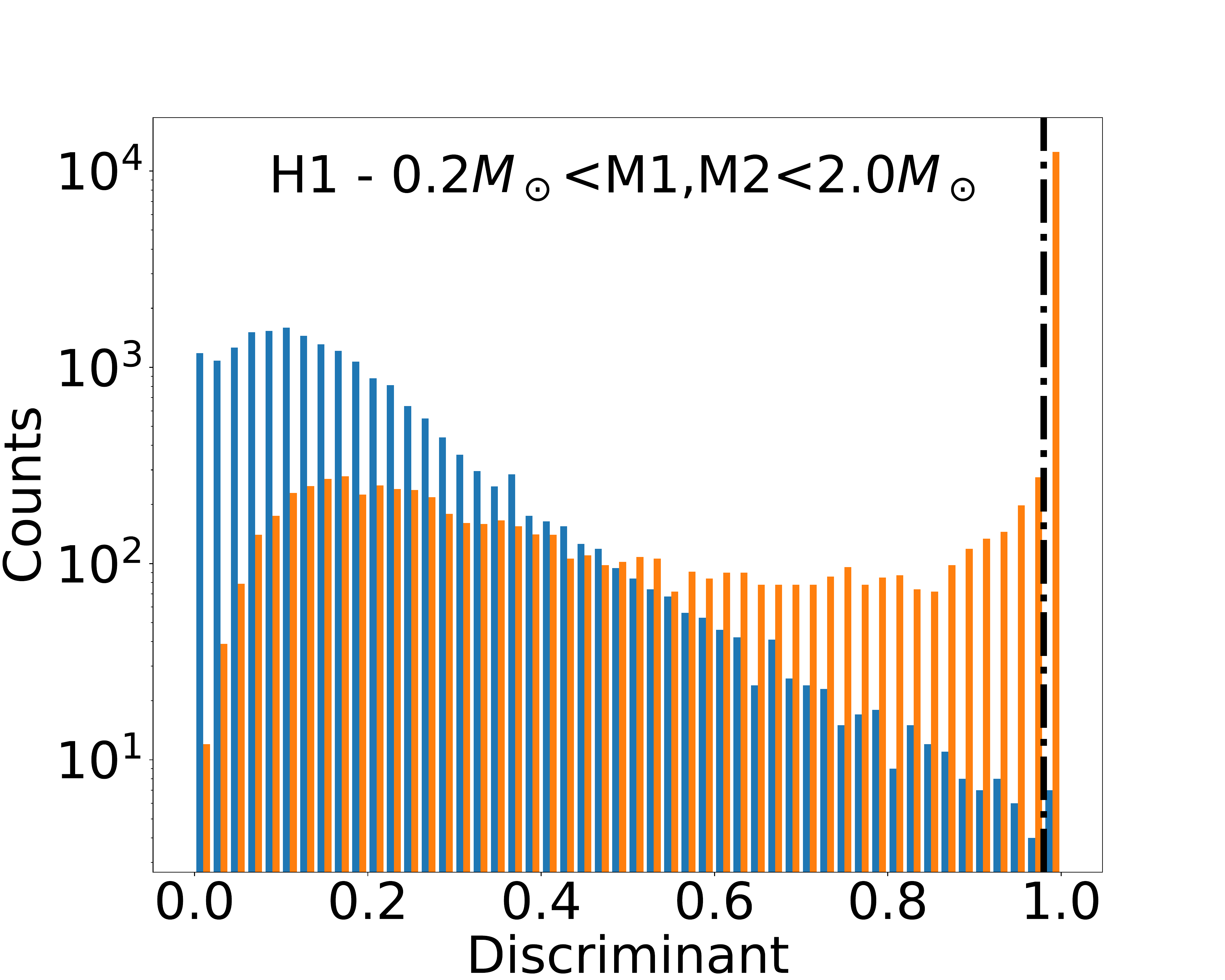}
\includegraphics[width=0.245\textwidth]{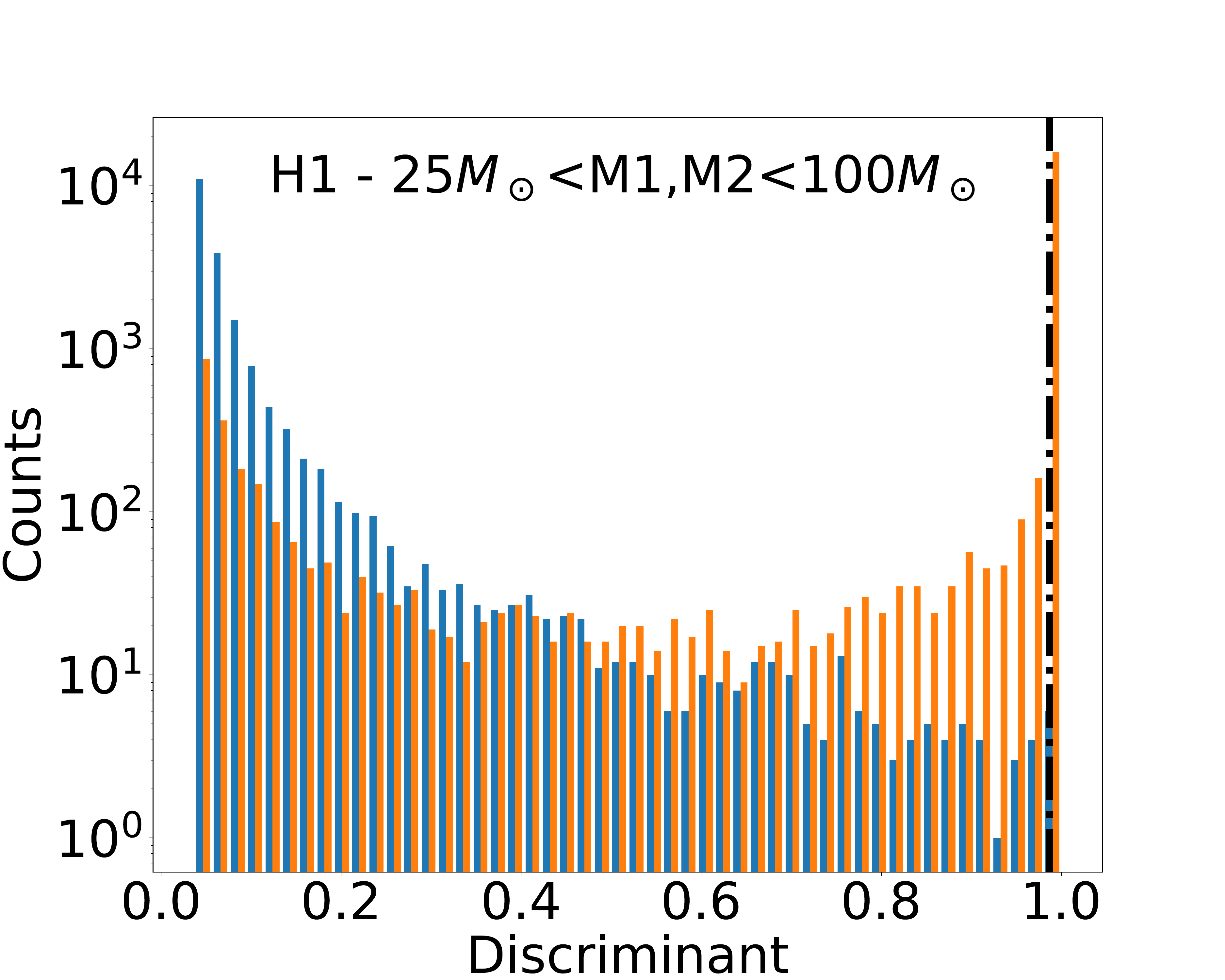}  
}
\\
\mbox{
\includegraphics[width=0.245\textwidth]{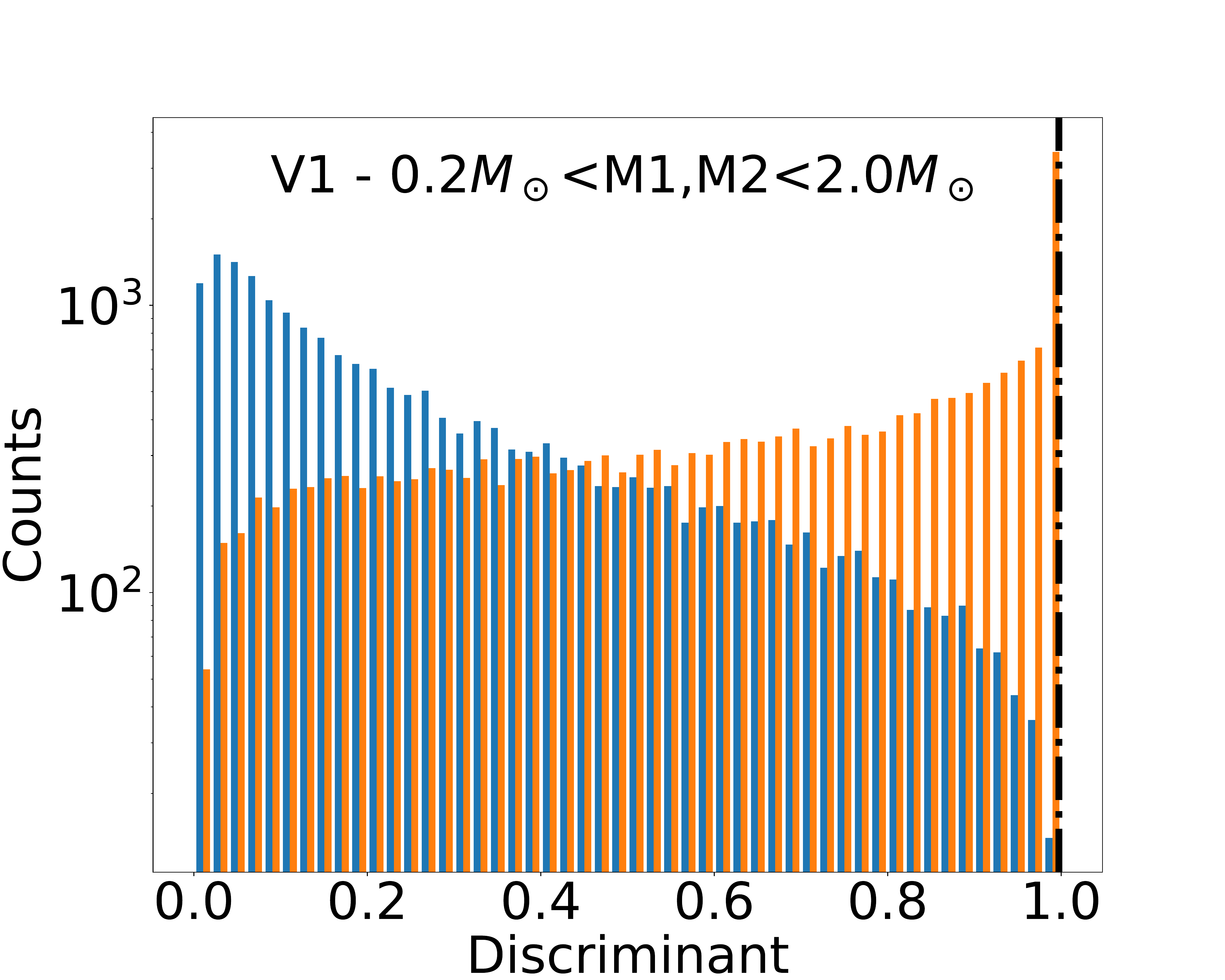}
\includegraphics[width=0.245\textwidth]{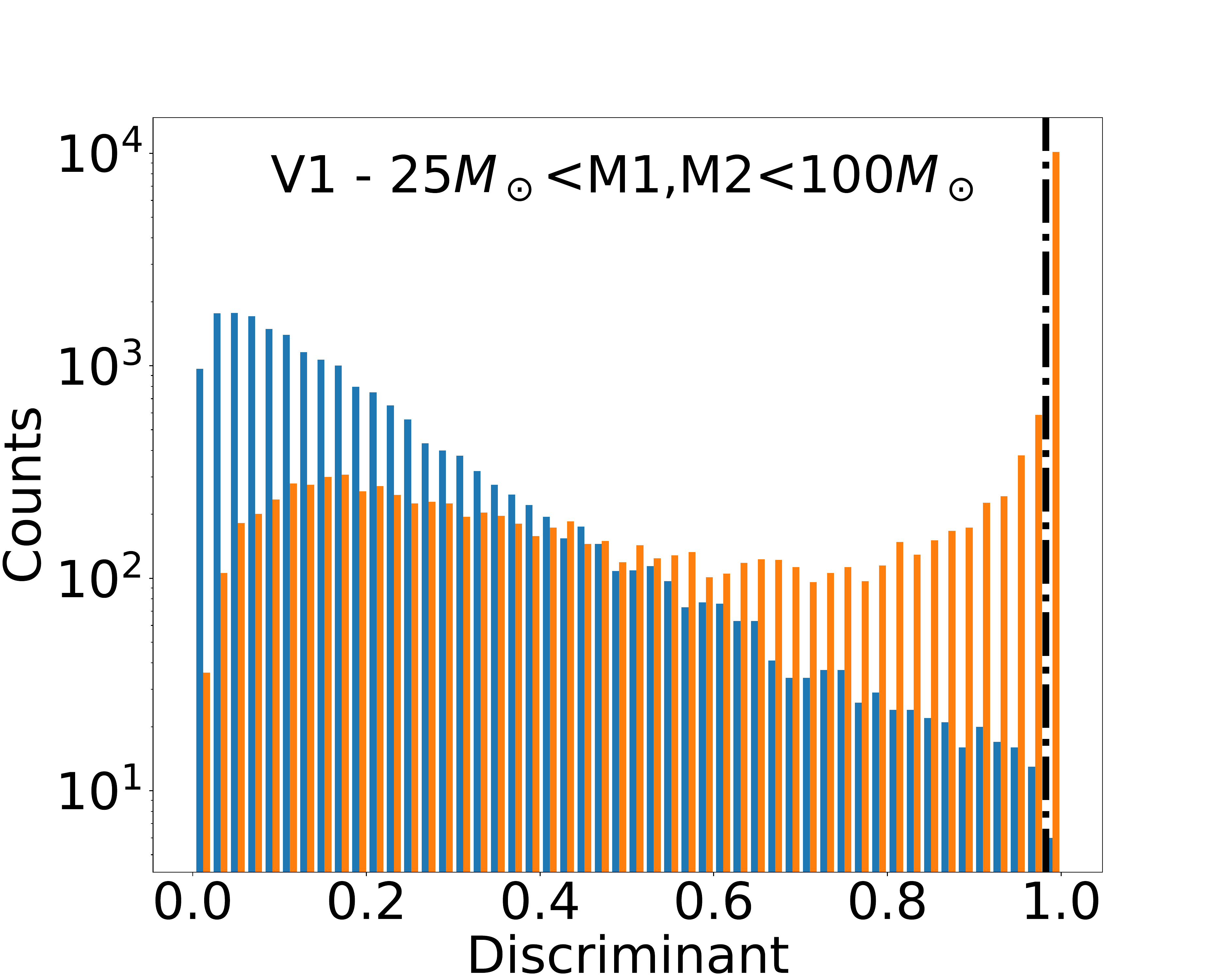}  
}
\end{center}
\caption{\small
CNN discriminating outputs corresponding to the (top) L1, 
(middle) H1, 
and  (bottom) V1 cases for background and signal images, 
in the case of (left) low mass  and (right) high mass ranges. 
The dashed-dotted lines indicate the thresholds used to identify the signal events.  
}
\label{fig:test}
\end{figure}

\begin{figure}[htb]
\begin{center}
\mbox{
\includegraphics[width=0.245\textwidth]{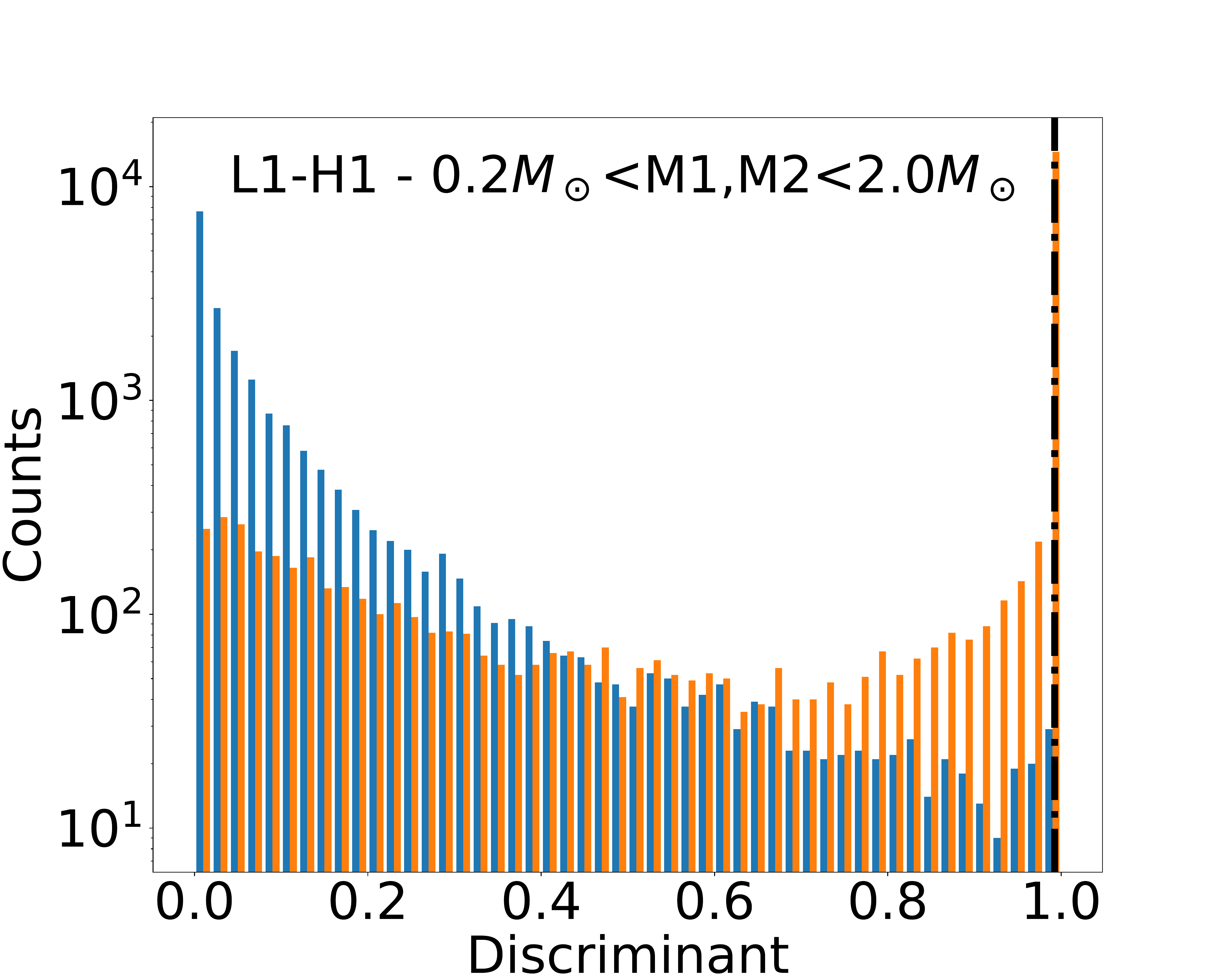}
\includegraphics[width=0.245\textwidth]{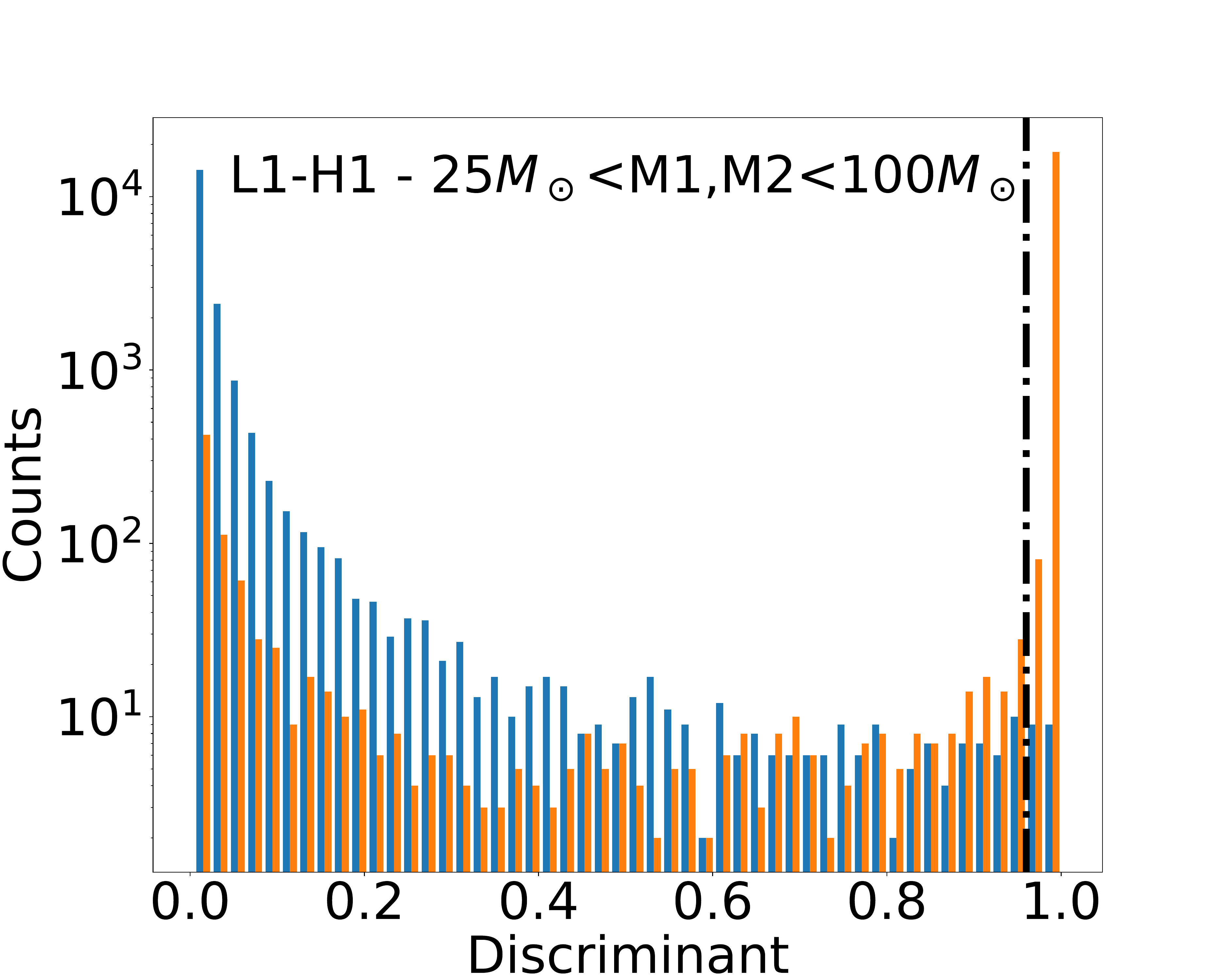}
}
\end{center}
\caption{\small
CNN discriminating outputs corresponding to the H1-L1  case for background and signal images
in the case of (left) low mass and (right) high mass ranges. The dashed-dotted lines indicate the threshold
used to identify the signal events.  
}
\label{fig:test2}
\end{figure}

Figures~\ref{fig:test} and ~\ref{fig:test2} present the final CNN output used for background and signal discrimination for the L1, H1 and V1, and the H1-L1 CNNs, respectively,  
as determined from the test samples, for both low mass  and high mass ranges.  The discriminant for the rest of the CNNs 
show similar features. 
A clear discrimination is obtained between signal and background samples.  
The ROC curves and the anticipated fake event rates 
are finally 
used as guidance in determining the final CNN threshold for classifying signal and background images. The CNN thresholds are 
adjusted such that the number of false positives in each case is limited to about 25 events per day. 
In Table~\ref{tab:nnth}, the information about the thresholds used in 
each case and the corresponding performance in terms of true positive and false positive rates  is collected.

\begin{table}[htb]
\begin{center}
\begin{footnotesize}
\begin{tabular}{l c c c} \hline
\multicolumn{4}{c}{single interferometer channel} \\ \hline
 & CNN discriminant ($\%$) & TP rate ($\%$) & FP rate ($\%$) \\ 
 & low/high mass & low/high mass & low/high mass \\ \hline
L1 & 98/96 & 70/92 & 0.09/0.06\\
H1 &98/99 & 65/84& 0.04/0.03 \\
V1 &99.7/98 &13/52 & 0.005/0.03 \\ \hline
%
% -----
%
\multicolumn{4}{c}{double interferometer channel} \\ \hline
 & CNN discriminant ($\%$) & TP rate ($\%$) & FP rate ($\%$) \\ 
 & low/high mass & low/high mass & low/high mass \\ \hline
L1 -- H1 & 99/96&74/95 &0.06/0.09 \\
L1 -- V1 &99/97 &69/93 &0.09/0.04 \\
H1 -- V1 & 99/98&66/88 &0.03/0.05 \\ \hline
\end{tabular}
\end{footnotesize}
\caption{\small 
Value of the CNN selection discriminant  in the low mass and high mass ranges,  
together with the anticipated true positive (TP) and false positive (FP) rates. 
}
\label{tab:nnth}
\end{center}
\end{table}

\section{Injection tests}

\begin{figure}[htb]
\begin{center}
\mbox{
\includegraphics[width=0.495\textwidth]{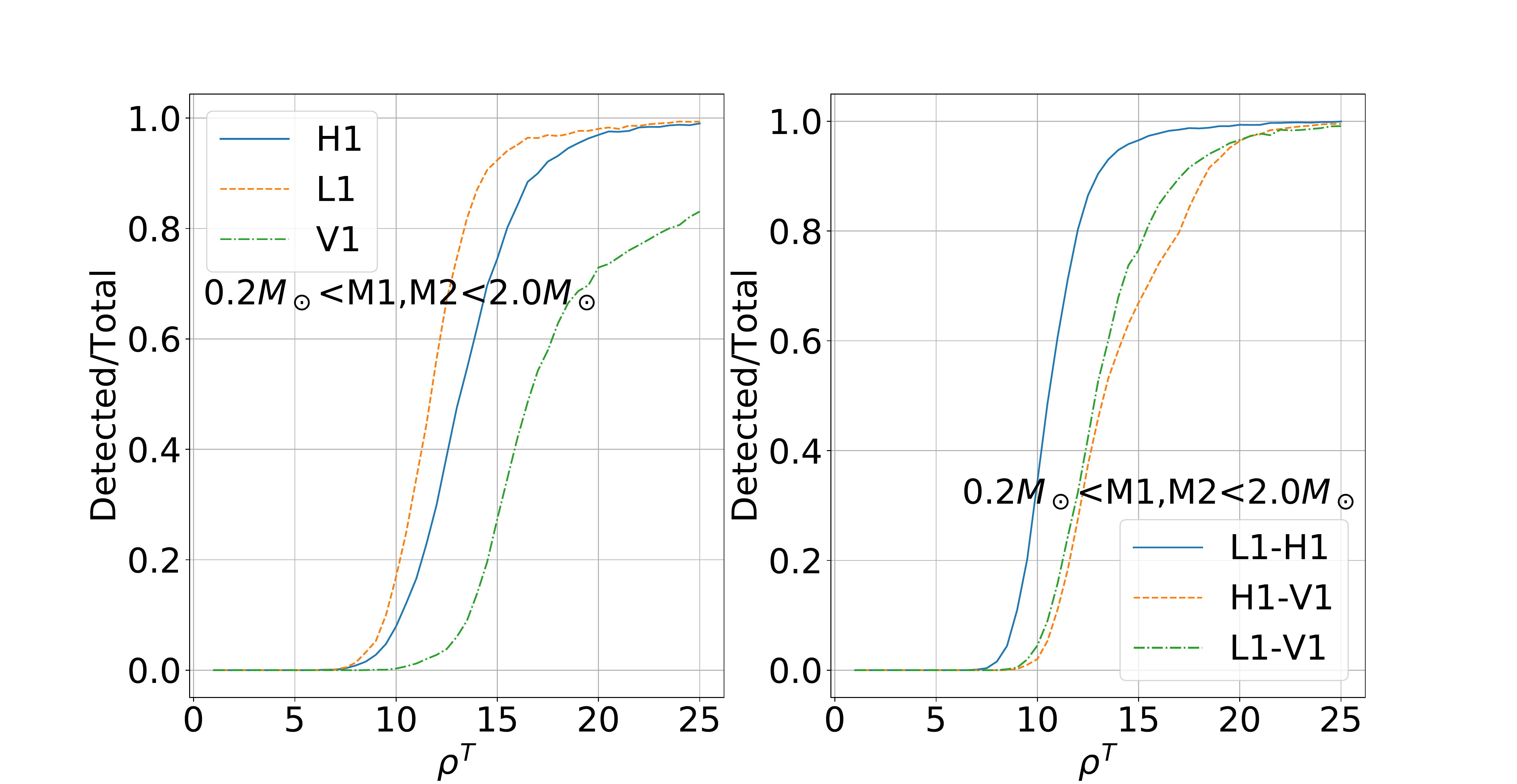}
}
\mbox{
\includegraphics[width=0.495\textwidth]{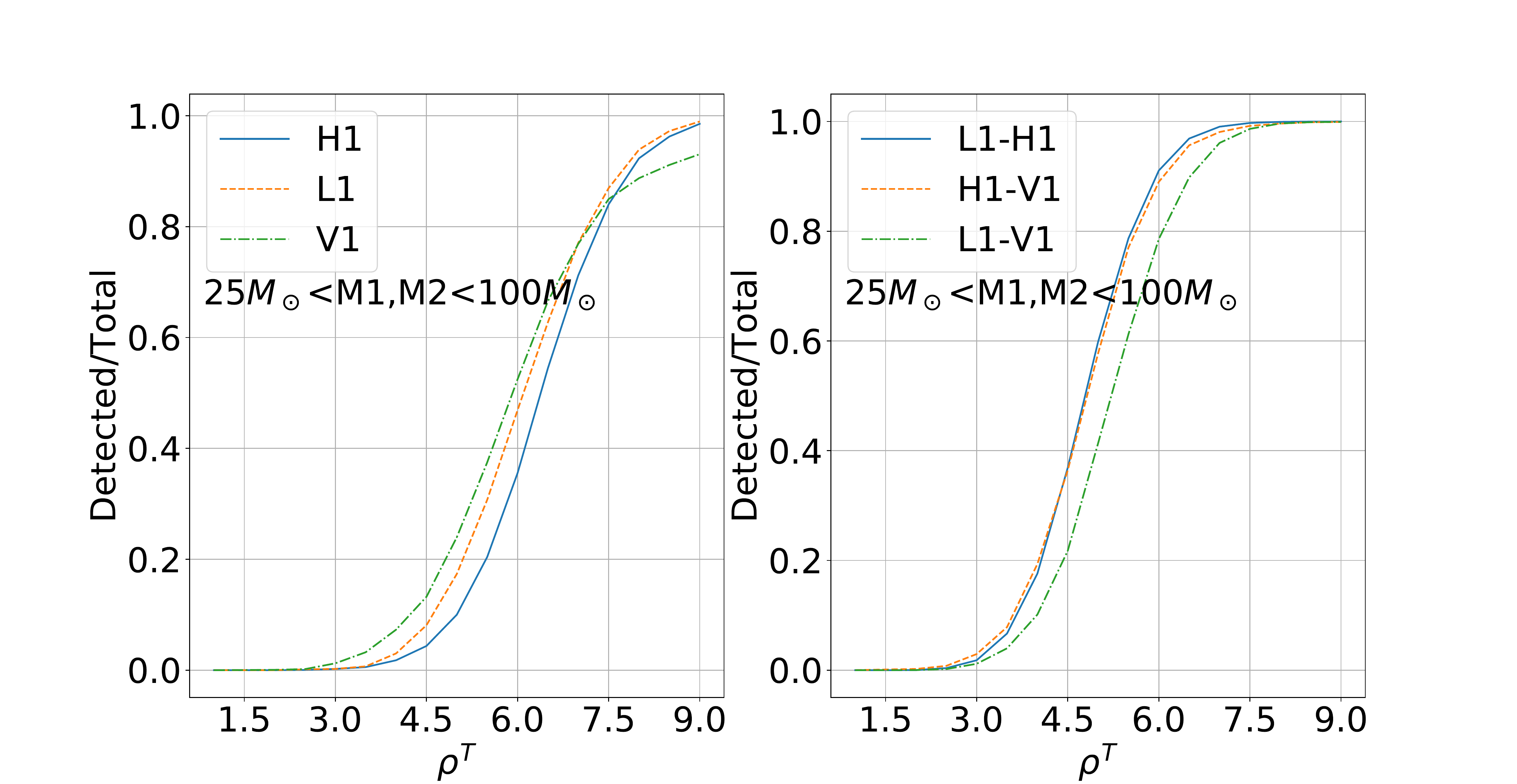}
}
\end{center}
\caption{\small
Efficiency vs $\rho^T$ for the different CNNs in the (top) low mass and (bottom) high mass ranges.
}
\label{fig:lowinject1}
\end{figure}

In order to determine the performance of the CNNs, injection studies of GW signals with given signal-to-noise ratios ($\rho$) are performed. 
The study is carried out separately for low and high mass ranges,  in which masses and distances are injected following a  homogeneous 
probability distribution. 
For each GW signal, the  value for $\rho$ is computed following the 
prescription in Ref.~\cite{PhysRevLett.120.141103} solving the integral 
\begin{equation}
    \rho^2 = \int_{f_{\rm{min}}}^{f_{\rm{max}}}  df \ |h(f)^2|/S_n(f), 
\end{equation}
\noindent
in the frequency domain $(f)$, where $|h(f)^2|$ denotes the signal and $S_n(f)$ 
is the power spectral density of the background. In the case of the background, a fixed period of time of 4096 seconds is used, whereas 
for the signal we employed the five-seconds window covered by the image. 
A Tukey window with $\alpha = 1/9$ is considered 
for the Fourier transformation.  The signal templates are then re-scaled  to targeted values 
$\rho^T$ by multiplying the amplitude of the signal 
by the ratio $\rho^T/\rho$.  
Figure~\ref{fig:lowinject1} shows the fraction of GW signals identified by the CNNs as a function of $\rho^T$   
in the case of inputs from single  and pairs of interferometers,  and for low- and high-mass ranges, respectively.  As expected, the efficiency for signal detection increases rapidly with 
$\rho^T$,  
becoming more efficient 
for large $\rho^T$ values.  
In the case of the high mass range,  the L1-H1 CNN provides the 
best results with an efficiency for selection of about 80$\%$ for $\rho^T = 6$,  becoming fully efficient for $\rho^T = 8$. The 
other CNNs involving Virgo performed differently but also become fully efficient around $\rho^T = 8$. In the case 
of the low mass range, the differences among CNNs become more evident,   with L1-H1 still providing the best performance 
with a 80$\%$ efficiency for $\rho^T = 12$, becoming fully efficient for $\rho^T = 16$.   Table~\ref{tab:rho} collects, separately for 
single interferometer  and pair of interferometers based CNNs, the values of  $\rho^T$ for given  signal detection efficiencies. 

\begin{table}[htb]
\begin{center}
\begin{footnotesize}
\begin{tabular}{l c c c} \hline
\multicolumn{4}{c}{single interferometer channel} \\ \hline
 & $\rho^T (50\%)$ & $\rho^T (80\%)$ & $\rho^T (99\%)$\\
 & low/high mass & low/high mass & low/high mass \\ \hline
L1 & 12.0/6.5& 13.5/7.5& 23.0/9.5  \\
H1 & 13.5/6.5& 15.5/7.5& 25.0/9.5 \\
V1 & 17.0/6.0& 23.5/7.5& $>$25/$>$20 \\ \hline
%
% -----
%
\multicolumn{4}{c}{double interferometer channel} \\ \hline
 & $\rho^T (50\%)$ & $\rho^T (80\%)$ & $\rho^T (99\%)$\\ 
 & low/high mass & low/high mass & low/high mass \\ \hline
L1 -- H1 & 11.0/5.0 & 12.0/6.0 &  19.0/7.0\\
L1 -- V1 & 13.0/5.5 & 15.5/6.5 &  24.5/8.0\\
H1 -- V1 & 13.5/5.0 & 17.5/6.0 &  23.0/7.5 \\ \hline
\end{tabular}
\end{footnotesize}
\caption{\small 
Values of $\rho^T$ at given detection efficiencies for the different 
CNNs using a  single detector or  detector pairs.
}
\label{tab:rho}
\end{center}
\end{table}

% ==============
% RESULTS
% ==============

\section{Results}
\label{sec:results}
The low mass  and high mass CNNs were used to search for candidate events in the O2 data. 
As shown in the previous sections,  the performance of the CNNs  that use  information 
from pairs of interferometers is slightly better than that from the CNNS relying on single detections, 
and therefore are used to obtain the final results. 
We first applied the CNNs discrimination to the data segments corresponding to the events 
included in the O1+O2 LIGO-Virgo catalog. 
The results are collected in Table~\ref{tab:O2ca}.
A majority of the events were properly identified by at least one of the CNNs.  
In particular, the GW170817 event, corresponding to the NS-NS event, 
was identified by the low-mass CNN,  
whereas the rest of the events in the catalog that were identified, 
corresponding to BH-BH events, triggered the high mass CNN.
All the rest of the events, 
for which none of the CNNs detected the signal,  
correspond to events with masses outside the ranges considered for the training.   

\begin{table}[htb]
\begin{center}
\begin{footnotesize}
\begin{tabular}{l| c c | c c } \hline
\multicolumn{5}{c}{CNNs response to O1+O2 catalog } \\ \hline
      & \multicolumn{2}{c|}{low mass } & \multicolumn{2}{c}{high mass} \\ \hline
Event & CNN & Detected & CNN  & Detected \\  
      &  value & (Y/N) & value   & (Y/N) \\ \hline
GW170104 &0.001 & N & 1.0& Y \\
GW170608 &0.02 &N &0.008 & N \\
GW170729 &0.1 &N & 1.0& Y \\ 
GW170809 & 0.15&N &1.0 & Y \\ 
GW170814 & 0.01&N &1.0 & Y \\ 
GW170817 & 1.0&Y &0.04 & N \\ 
GW170818 & 0.003&N & 1.0& Y \\ 
GW170823 & 0.05&N &1.0 & Y \\
GW150914 (O1) & 0.24&N &1.0 & Y \\
GW151012 (O1) & 0.06&N &0.95 & N \\
GW151226 (O1) & 0.29&N &0.08 & N \\ \hline
\end{tabular}
\end{footnotesize}
\caption{\small
Summary of the CNN response to the O1+O2 catalog events. 
}
\label{tab:O2ca}
\end{center}
\end{table}

\begin{figure}[htb]
\begin{center}
\mbox{
\includegraphics[width=0.245\textwidth]{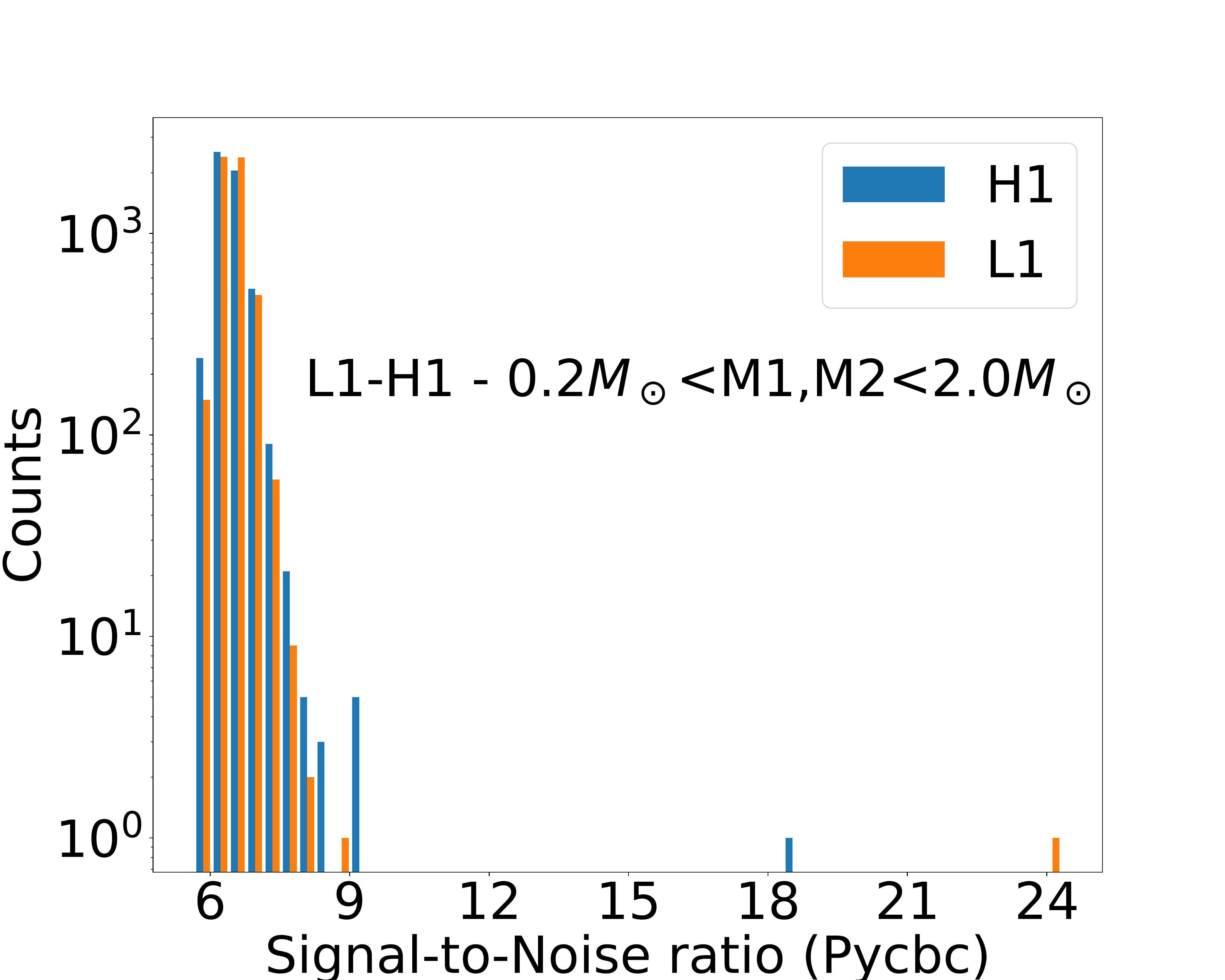}
\includegraphics[width=0.245\textwidth]{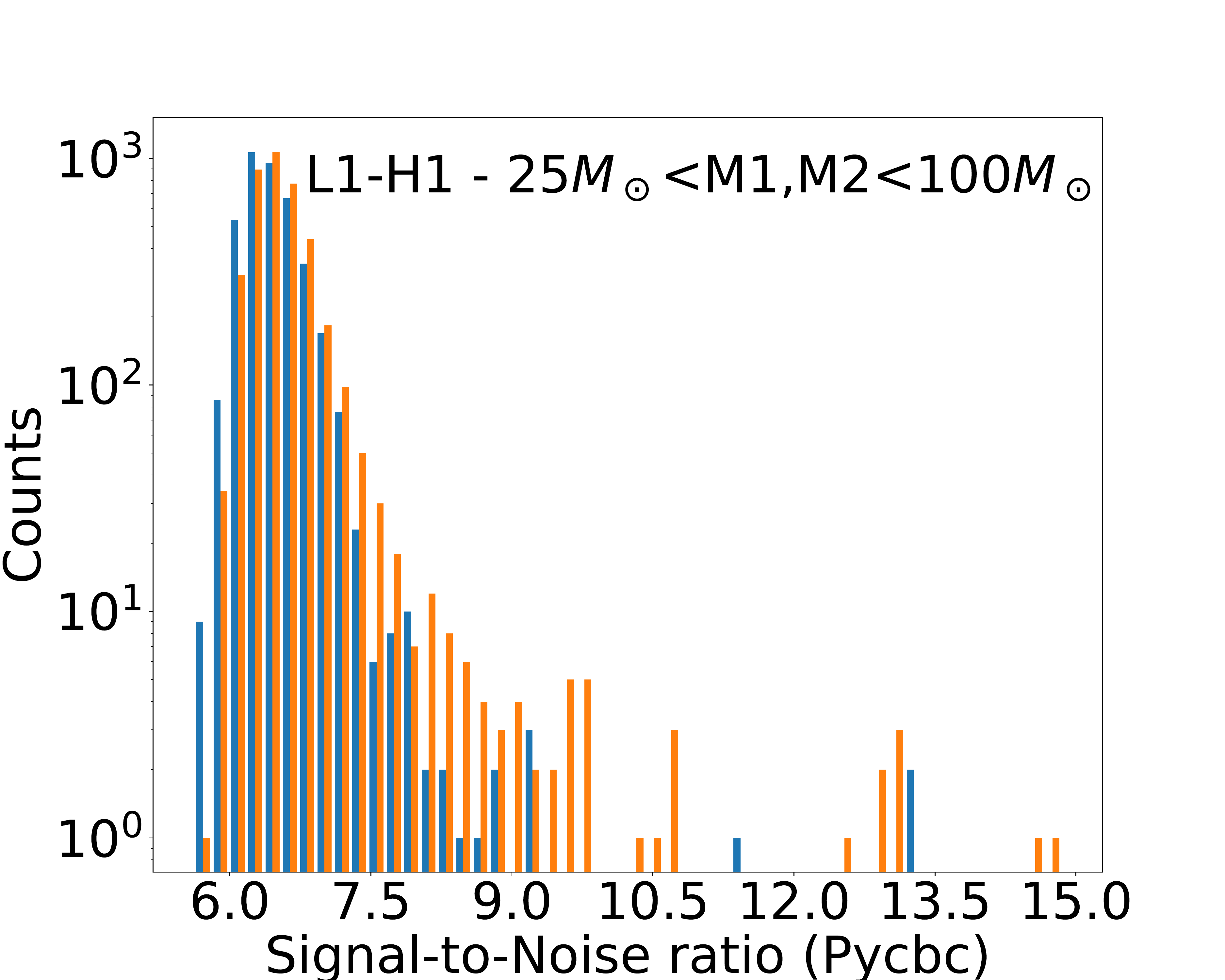}
}
\end{center}
\caption{\small
Distribution of signal-to-noise ratio values, as determined using pyCBC pipeline, 
for the events detected by the  (left) low mass and  (right) high mass CNNs.  
}
\label{fig:SNR}
\end{figure}

We then performed a  scan of the full O2 data set, 
for which a slicing window of five seconds duration was used in steps of 2.5 seconds 
(leading to a 50$\%$ overlap between consecutive images) in each of the interferometers.  
In the case of the L1-H1, 
more than four million images are tested for the presence of potential signals. 
For H1-V1 and L1-V1, 
data are reduced to about 584 thousand and 601 thousand images, respectively, 
as dictated by the limited duration of the Virgo data set. 
The results of the low mass and high mass CNN scans are collected in Table~\ref{tab:O2full}. 
The L1-H1 CNNs  identify about 0.1$\%$ of the images as potential signals,
corresponding to an average detection rate of about 47 and 34 events per day for the low mass and the high mass CNN, 
respectively.  
In the case of the H1-V1 CNNs, 
only $0.07\%$ of the images are triggered as potential signals, 
leading to an average detection rate of 25 events  per day.  
The performance of the  high mass L1-V1 CNN is very similar to that of the H1-V1 case, 
whereas the performance of the low mass L1-V1 CNN is significantly worse, 
triggering on $0.5\%$ of the images and with an average detection rate of almost 180 events per day.
As shown in Table~\ref{tab:O2full}, 
the rate of detection is large compared to the size of the O2 catalog, and the quoted daily detection rates present 
large variations over time (with up to a $50\%$ r.m.s), 
indicating that the CNNs are triggering on noisy events. 
However, no attempt is made to re-tune the CNNs  since no abrupt systematic increase of triggers was observed over time.
In order to determine the significance of the different detections in terms of signal-to-noise ratio,  
the ensemble of images triggered by the low mass and high mass CNNs are correlated to an independent 
scan over the O2 available single trigger data~\cite{singleTriggerData} using the pyCBC pipeline~\cite{Pycbc}. 
A large majority of the images flagged as potential signals by the CNNs are also detected by the pyCBC pipeline with signal-to-noise ratio in the range between six and eight (see Figure~\ref{fig:SNR}). 

This indicates that the CNNs are performing similarly to the first steps of the dedicated pipelines.  
Once consistency across interferometer signals is imposed in the pyCBC single-trigger list, 
the O2 catalog candidates emerge. 
In the case of the high-mass CNN, 
three events (less than 0.08$\%$ of the total sample) 
were flagged by the CNN with no pyCBC corresponding detection. 
Similarly, 
six events (less than 0.1$\%$ of the total sample) 
were detected by the low-mass CNN without having a corresponding detection by the pyCBC pipeline.   
After a visual inspection, 
the nine events were checked to be due to noise,
with no coincident signals in both L1 and H1 intereferometers, 
pointing to low (under threshold) signal-to-noise values.

\begin{table}[htb]
\begin{center}
\begin{footnotesize}
\begin{tabular}{l| c  c c | c c } \hline
\multicolumn{6}{c}{CNNs Response to full O2 scan} \\ \hline
      & & \multicolumn{2}{c|}{low mass NN} & \multicolumn{2}{c}{high mass NN} \\ \hline
CNN  & Images & Detected & Events/day & Detected & Events/day \\ \hline 
L1-H1  & 4077233 & 5496 & 47 &  3973 & 34 \\
H1-V1  &584993 & 439 & 26 & 414 & 24 \\
L1 -V1 &601877 & 3078 & 178 &445 & 26\\ \hline
\end{tabular}
\end{footnotesize}
\caption{\small
Results of the low mas and high mass scans over the full O2 data for the different CNNs considered.  
The number of images processed and detected are reported together with the  corresponding average 
daily detection rate. 
}
\label{tab:O2full}
\end{center}
\end{table}

% ==============
% SUMMARY
% ==============

\section{Summary}
\label{sec:sum}
We have presented the result of studies using convoluted neural networks based on the ResNet50 architecture 
to search for compact binary coalescence of black holes in the LIGO-Virgo data from the O2 observation run. 
Two separate CNNs are trained specifically for low mass ($0.2 - 2 \msun$) and high mass ($25 - 100 \msun$) black holes,  
and the training process explores the simultaneous use of pairs of interferometers as input.  
The simultaneous use of the three interferometers in the CNN was not allowed by the limited size of the Virgo O2 data.  
A performance comparable to that of the first detection steps of the dedicated pipelines
is reported for the CNNs in terms of efficiency and purity in selecting signal events. 
A scan over the full O2 data set is carried out demonstrating that the CNN 
response is similar to that obtained in matched-filtering based pipelines. 
All the O1+O2 catalog events, 
with masses compatible with the training parameters, 
are identified by the CNNs,  
and no new events are detected with a significant signal-to-noise ratio. 
This study shows the viability of CNN-based pipelines and could be regarded as a step towards an online implementation, 
in preparation for the future LIGO-Virgo-KAGRA combined  observation runs. 
Future studies will extend the CNN training towards the simultaneous use of multiple interferometers relevant for O3 and O4 observation runs. 

% ======================
% Acknowledgments
% ======================
\section*{Acknowledgements}
The authors would like to thank M. Eriksen, E. Cuoco and M. Razzano for their helpful comments and discussions. The authors are grateful for computational resources provided by PIC (Spain) and by the LIGO Laboratory and supported by National Science Foundation Grants PHY-0757058 and PHY-0823459. This paper has been given LIGO DCC number LIGO-P2000521.  This work is partially  supported   by  the Spanish MINECO   under
the grants SEV-2016-0588 and PGC2018-101858-B-I00, some of which include
ERDF  funds  from  the  European  Union. IFAE  is  partially funded by
the CERCA program of the Generalitat de Catalunya. This work was carried out within the framework of  the EU COST action CA17137. 

% ==============
% BIBILIORGAPHY
% ==============

\bibliographystyle{apsrev}
\bibliography{mybib}{}

\end{document}